# Anti-microbial properties of a multi-component alloy

Anne F. Murray[1,2], Daniel Bryan[1], David A. Garfinkel[3], Cameron S. Jogensen[3], Nan Tang[3], WLNC Liyanage[3], Eric A. Lass[3], Ying Yang[4], Philip D. Rack[3], Thomas G. Denes[1], and Dustin A. Gilbert[3,5*]

[1] Department of Food Science, University of Tennessee, Knoxville, Tennessee 37996
[2] Department of Ecology and Evolutionary Biology, University of Tennessee, Knoxville, Tennessee 37996
[3] Department of Material Science, University of Tennessee, Knoxville, Tennessee 37996
[4] Oakridge National Laboratory, Material Science and Technology Division, Oakridge, Tennessee, 37831
[5] Department of Physics and Astronomy, University of Tennessee, Knoxville, Tennessee 37996
* Corresponding author (D.A.G.): E-mail: dagilbert@utlk.edu

**Abstract**

High traffic touch surfaces such as doorknobs, countertops, and handrails can be transmission points for the spread of pathogens, emphasizing the need to develop materials that actively self-sanitize. Metals are frequently used for these surfaces due to their durability, but many metals also possess antimicrobial properties which function through a variety of mechanisms. This work investigates metallic alloys comprised of several bioactive metals with the target of achieving broad-spectrum, rapid bioactivity through synergistic activity. An entropy-motivated stabilization paradigm is proposed to prepare scalable alloys of copper, silver, nickel and cobalt. Using combinatorial sputtering, thin-film alloys were prepared on 100 mm wafers with ≈50% compositional grading of each element across the wafer. The films were then annealed and investigated for alloy stability. Bioactivity testing was performed on both the as-grown alloys and the annealed films using four microorganisms – Phi6, MS2, *Bacillus subtilis* and *Escherichia coli* – as surrogates for human viral and bacterial pathogens. Testing showed that after 30 s of contact with some of the test alloys, Phi6, an enveloped, single-stranded RNA bacteriophage that serves as a SARS-CoV 2 surrogate, was reduced up to 6.9 orders of magnitude (>99.9999%). Additionally, the non-enveloped, double-stranded DNA bacteriophage MS2, and the Gram-negative *E. coli* and Gram-positive *B. subtilis* bacterial strains showed a 5.0, 6.4, and 5.7 log reduction in activity after 30, 20 and 10 minutes, respectively. Bioactivity in the alloy samples showed a strong dependence on the composition, with the log reduction scaling directly with the Cu content. Concentration of Cu by phase separation after annealing improved activity in some of the samples. The results motivate a variety of themes which can be leveraged to design ideal bioactive surfaces.

**Introduction**

Touch surfaces in high-traffic areas[1] can become vectors for disease propagation through indirect contact between infected and vulnerable persons,[2, 3] making it critical to develop self-sanitizing materials which are effective against a broad range of pathogens. Previous works have shown that microorganisms can remain alive or active on surfaces for hours to days,[4-7] including many human pathogens such as Methicillin susceptible *Staphylococcus aureus* (MSSA) and resistant *Staphylococcus aureus* (MRSA),[8] Rhinovirus[9], Influenza virus A[10], Rotovirus[11] and corona viruses such as the severe acute respiratory syndrome coronavirus 2 (SARS-CoV-2),[5] which has caused the global pandemic of 2020-2022.[12] These long lifetimes partly determine the potential for pathogens to spread during subsequent contact with the surface. High-traffic surfaces are often made of metals, with stainless steel (SS) being a common choice due to its relatively low cost, durability, and resistance to corrosion; SS has been reported to have modest anti-microbial properties.[13] Occasionally brass (CuZn), an alloy of copper and zinc, is employed for touch surfaces due to its pleasing golden hue, however it is a much softer, more expensive – three times the cost of SS at the time of writing – and readily oxidizes; the green or black oxidation tends to exfoliate from the surface and so it is regularly cleaned if used as a touch surface. Brass has known anti-microbial activity. [14, 15] Focusing on disease propagation, many of the bioactive metals[16-24] such as copper (Cu), silver (Ag), and cobalt (Co), are not widely employed due to their cost and tendency to oxidize. Each of these metals achieves bioactivity by different mechanisms against diverse pathogens. A solid-solution alloy of these elements may be active against a range of pathogens larger than the sum of its parts utilizing a combination of anti-microbial mechanisms, resulting in a bioactive super alloy.

This work focuses on the development of multicomponent alloys of CuAgCo; each of these metals achieve bioactivity through different modes of action.[24] Copper in-particular has shown biological activity against a wide range of viruses[5, 16-18, 25] and bacteria.[5, 24] The mode of action in these systems has been attributed to interactions between the $Cu^{+1/+2}$ ions[26] and surface proteins which become denatured, resulting in the viral envelope failing.[27, 28] Silver metal has been used as an anti-microbial agent since times of antiquity[29] and can bind with virus surface glycoproteins disrupting replication.[19, 20, 30, 31] In bacteria, the sanitizing mechanisms in Ag have been attributed to damage to the cell wall and membranes[32] and interference with internal cellular functions.[24, 33] Cobalt in the $Co^{3+}$ state has been reported to have anti-bacterial and anti-viral[21-23] properties when complexed with chelators or ligands, potentially through Schiff bases, a mechanism that inactivates protein active sites.[34] In its un-oxidized state, cobalt has been shown to be effective at reducing bacterial presence.[33, 35] By developing an alloy of these metals, the resultant material may be bioactive through a range of mechanisms, making it simultaneously effective against a range of pathogens larger than any one metal. Furthermore, the multifaceted modes of action may provide accelerated sanitation properties.

A challenge to this work is that CoCu, AgCo, and AgCu do not mix to form alloys, e.g. they are immiscible.[36] To overcome this challenge, two approaches are used. In the first, room-temperature sputtering is used to achieve rapid solidification, freezing the atoms as an alloy before they are able to phase separate.[37, 38] In the second approach, an entropy-based stabilization paradigm[39, 40] will be leveraged. This materials design concept proposes that increasing the chemical entropy within a multicomponent system increases the free energy cost of phase separation and thus can stabilize alloys of elements which are typically immiscible.[41] This approach has been used previously to realize new functional materials such as catalysts[42] and magnets[43, 44]. To increase the chemical entropy, an additional element, nickel (Ni), is added to form the four-component alloy CoCuAgNi; nickel does possess some bioactivity of its own[45] but was chosen primarily because it is miscible with both Cu and Co across their entire compositional range. In Ni-rich ternary compositions, CoCuNi forms a single-phase alloy. During the annealing step silicon (Si) is also introduced, diffusing from the substrate. Attempts were also made to include Iron (Fe) and Palladium (Pd), forming the five-component alloy CoCuAgFePd. A recent investigation has shown potential bioactivity of a copper-containing high-entropy alloy (HEA) on two viruses.[46]

In the present study, thick-films of a proposed bioactive HEA CuAgCoNi(Si) are prepared by combinatorial sputtering and tested for their potential application as self-sanitizing surfaces in high-traffic environments. The bioactivity of the as-prepared alloy is tested against four non-pathogenic surrogates representing infectious human pathogens including the bacteriophage, Pseudomonas phage Phi6, which is similar in size and structure to SARS-CoV-2, including the surface spike proteins,[47] and the phage MS2, a non-encapsulated spike-less virus and a surrogate for the human novovirus.[48] Bacterial testing was performed on the Gram-positive *Bacillus subtilis* NRS 231 (ATCC: 6633) and the Gram-negative *Escherichia coli* strain Seattle 1946 (ATCC: 25922), which served as surrogates for pathogenic bacterial strains. Next, the films were annealed and the bioactivity re-examined. For both the as-grown and annealed films, the nano- and microstructure is evaluated to test the HEA paradigm and assess the structure-performance relationship. In addition to the sputtered films, a range of related singe-component, binary and ternary surfaces are also investigated. This approach is designed to address the critical question of how the nanoscopic distribution of elements within a surface affects its bioactivity.

**Results**
*Structure of the sputtered thin films*

Thin-films of (NiAgCoCu) alloy were prepared by combinatorial sputtering at room temperature on a 100 mm Si (100) wafer, as described in the Methods section. The sputtering technique uses an accelerated Ar plasma to generate a vapor of each metal at single-element sources. The elements blend in the vapor state and condense on the Si substrate, with solidification occurring on the nanosecond timeframe, offering little opportunity for the elements to diffuse. Without the opportunity to diffuse, the elements can form a metastable alloy, even if the parent elements are immiscible.[37, 38] Evidence of alloying, rather than amorphous or glassy behavior, can be seen in the X-ray diffraction (XRD) data discussed below. For combinatorial sputtering specifically, the sputtering sources are arranged around the substrate in a confocal geometry; regions of the wafer close to (far from) a source have a higher (lower) concentration of the element from that target, thus a compositional gradient of ≈50% is achieved across the diameter of the wafer.

Compositional mapping was performed by scanning electron microscopy (SEM) with energy dispersive X-ray spectroscopy (EDX), measured at 17 positions on the wafer. These maps, shown in Figure 1a, identify the compositions at the four extremum edges of the sample to be ($Ni_{11}Ag_{21}Co_{13}Cu_{55}$) ($Ni_{22}Ag_{54}Co_{6}Cu_{18}$) ($Ni_{26}Ag_{12}Co_{39}Cu_{23}$) ($Ni_{55}Ag_{22}Co_{12}Cu_{11}$), with an equal compositional point located near the wafer center. X-ray diffraction measurements of the as-grown sample, Figure 1b, show broad peaks at ≈39° (d=2.31 Å) and ≈44° (d=2.06 Å); these peaks align closely with the parent compounds: face centered cubic (FCC) Ag (111) at 37.4° (d=2.40 Å), Ag (200) at 43.5° (d=2.08 Å) and FCC Ni (111) at 44.8° (d=2.02 Å). The similarity between these parameters suggests that the as-grown films possess an FCC structure with a polycrystalline microstructure. The data suggests that, for sufficient Ag content, the smaller (Cu, Ni, Co) atoms are integrated into the Ag crystal lattice. Decreasing the Ag content, the peak at ≈39° disappears, while the peak at ≈44° moves to higher angles, indicating the Ag is now integrated into the smaller Ni lattice. The system is inherently strained due to the large difference in atomic size between the Ag (diameter of 172 pm), and the Ni, Co and Cu (≈130 pm)[49]. A crude measure of the structural phase distribution can be inferred from the ratio of the XRD peak areas, after correcting for the Fresnel fall off, the overlap of the Ag(200) and Ni(111), and the larger form factor from the Ag atoms. The resultant map of the phase distribution, Figure 1c, confirms that the system undergoes a structural transition from the larger to smaller lattice spacing and at the equal composition point, the Ag is largely integrated into the NiCoCu lattice.

After annealing at 600 °C for 60 minutes in flowing forming gas ($N_2$ + 5%$H_2$) the samples developed a foggy surface contrast while retaining a grey metallic hue, suggesting the they were not oxidized; annealing at 600° then cooling in air resulted in a blue colored oxide. X-ray diffraction measurements of the annealed films are shown in Figure 2 and are indexed to three primary phases: Ag ($FCC^1$), (Ni,Co)$Si_2$ (fluorite, $FCC^2$), and (Ni,Co,Cu)Si (B20). Previous works have shown that CoNi films grown on Si substrates and annealed in similar conditions uptake Si from the substrate, forming these phases.[50, 51] Identification of these particular compositions were informed by previous works on NiCo/Si systems, and by comparison between the XRD and EDX results. All of the XRD patterns show peaks for an Ag-like ($FCC^1$) phase with a lattice parameter of 4.077 Å, matching closely to bulk Ag (4.20 Å). EDX images in Figure 3, taken at the equal composition point (identified as position 3 in the diagram) show ellipsoidal Ag precipitates ≈600 nm in diameter fully ejected from the film. The EDX also shows Ni and Co in the precipitates. The slightly smaller lattice parameter of FCC[1] compared to bulk Ag is likely the result of a small amount of residual Ni and Co being alloyed into the lattice.

The film beneath the Ag precipitates is comprised of two distinct domains, identified by EDX as Cu-rich and Cu-poor; Cu-rich regions are also identified as Si-poor when compared to the Cu-poor domains. Initial considerations were that these were (Ni,Co) and (Ni,Co,Cu) alloys,[36, 52] since the Si signal can also be from the substrate. However, the XRD results cannot be fitted using NiCoCu alloys, which are known to have a lattice parameter of ≈3.52 Å.[36] Previous works[50, 51] have shown that annealing NiCo on a Si wafer can rapidly form $NiSi_2$, CoSi, and $CoSi_2$, with heats of formation of -50.73 kJ/mol, -56.00 kJ/mol, and -40.7 kJ/mol respectively. The current work is further complicated by the presence of Cu. The XRD peak at 47° can be exclusively associated with the (220) peak of a $NiSi_2$ fluorite structure, hereafter identified as FCC[2], corresponding to a lattice parameter of 5.42 Å; this value is in good agreement with the parent compound, 5.44 Å. Similarly, the XRD peak at 50° can be associated exclusively with the (211) peak of a CoSi B20 structure, corresponding to a lattice parameter of 4.49 Å; this value is in good agreement with the parent compound, 4.43 Å. Using the diffraction peaks at 47° (red) and 50° (blue) the relative phase distributions are plotted in Figure 3. The phase map confirms that, near the Ni edge the film is single-phase FCC[2] ($NiSi_2$), while in the NiCo corner, the film is single-phase B20 (CoSi). This phase distribution also gives rise to a distribution of the electronic structure of the film, with the disilicate FCC[2] phase being predominantly charge neutral,[53] while the monosilicate B20 phase is tetravalent.

Recognizing that the Cu-rich and Si-poor regions are commensurate in the EDX images, the Cu is proposed to preferentially incorporate into the monosilicate B20 structure. Based on these results, the annealed sample can be generally described as a film, with domains of $(Ni,Co)Si_2$ and (Ni,Co,Cu)Si, with (Ni,Co):Ag precipitates. The phase distribution map shows large, single-phase regions near positions 8 (B20) and 17 (FCC[2]). EDX images from position 8 show some chemical phase separation of the Cu, albeit less well-defined than position 3, despite that the XRD does not show structural phase separation.

Interesting microstructural changes occur approaching the extremums of the annealed wafer. Approaching the CoCu edge (position 14), the Ag does not precipitate as particles, but remains as part of the film, as a new domain. Approaching the Ag and AgCu edges (positions 1 and 6), the microstructure changes significantly, forming a worm-like lamella network suggesting at higher temperatures a single phase may exist and undergoes spinodal decomposition into Ag- and Cu-rich phases. The lamellae are much smaller than the domains or precipitates in the rest of the film, with widths of ≈65 nm. The different structures, particularly of the Ag, may result in different strengths of bioactivity.[54] These results suggest that Ag-rich and CoCu-rich compositions may be prepared as bulk alloys, and retain all of the elements in localized alloy domains. The following sections investigates the bioactivity of the as-grown wafer (with the elements uniformly intermixed on an atomic scale), and the annealed wafer (with compositional domains and Ag precipitates).

*Bioactivity Testing*

*Single-component and binary alloys*

A range of single-component metals, including Co, Cu, Ni and Ag, among others, were tested for bioactivity against the surrogate microbes. Metals for the single-component tests were included due to their reported bioactivity against a variety of human pathogens[24] or commercial availability. Prior to testing, surfaces were cleaned with acetone to remove any contaminating microbes or grease, but leaving the native oxide layer which would be expected on a deployed touch surface. Testing was performed by spotting 10 µL of a concentrated aqueous solution containing the test organism onto a 25×25 mm$^2$ test surface, covering the spot with a 25x25 mm$^2$ sterile glass slide, then recovering the microorganisms by washing the surface, and enumerating the live microorganisms, as described in the Methods. This testing methodology is designed to simulate wet contact, for example, microorganisms in droplets from breathing, sneezing or coughing and does not represent 'dry' contact assays. The metal Cu had significant ($p>0.05$) bioactivity against Phi6 and MS2, Figure 4a and 4b (and Supplemental Figure 1), achieving a 6.9-log reduction of Phi6 (30 second test time) and a 7.1-log reduction for MS2 (30-minute test time). To allow accurate comparisons, the data is normalized to the titer concentrations for each test. All of the other tested metals – including our candidate metals Co, Ni and Ag – showed much lower activity. Also notable among the tested metals is zinc (Zn), which has been previously reported to have bioactivity against a wide range of pathogens,[24] but here appears to be weak (Supplementary Figure 2). Zinc was not used in the wafer due to its high vapor pressure, making it incompatible with general-use vacuum systems. All of the tested surfaces, other than Cu, are statistically identical to the stainless steel (alloy 304) control; 304SS is chosen as the control due to its common use for high-traffic surfaces.

The gram-negative model *Escherichia coli* and gram-positive model *Bacillus subtilis* were also tested on the single-component surfaces, Figure 4c and 4d, respectively. These measurements again show strong activity from Cu, a significant 5.9-log reduction for both microorganisms after 20 minutes and no detectable *B. subtilis* after 10 minutes. No *B. subtilis* was recovered on any Cu control surface throughout this study. In all of the bacterial assays the Ag, Co and Ni were again statistically indistinguishable from the SS control. These results suggest that surface contact of these bulk materials may be insufficient to impart bioactivity in the timeframe tested here; testing times were determined by the Cu control. Overall,

Cu was the most effective metal at reducing microbial activity in both the phages and bacteria. This strong activity motivated its use as the positive control in subsequent assays.

Two alloys of brass (alloy 360, $Cu_{0.6}Zn_{0.4}$ and alloy 260, $Cu_{0.7}Zn_{0.3}$) were also examined in contact assays; these alloys contained both bioactive materials Cu and Zn (Supplementary Figure 3). Assays performed with Phi6 on alloy 260 and 360 resulted in no countable plaques after 30 seconds of exposure, identical to pure Cu. For the other phage, MS2, the bioactivity was superior when compared to any of the single-component surfaces except pure Cu; alloy 260 achieved a log reduction of 5.3, while alloy 360 achieved a log-reduction of 4.0. Assays performed on the bacteria showed a similar trend, with the Cu-rich alloys displaying strong bioactivity: *E. coli* showed a log-reduction of 5.6 on alloy 260 and no bacteria were recovered on alloy 360, while *B. subtilis* experienced a log-reduction of 4.5 and 5.2 on alloys 260 and 360, respectively.

*Copper Oxides*

The above results together support a theme that Cu-content strongly influences the bioactivity in contact assays. In a deployed surface, the Cu is expected to be oxidized and present as either divalent CuO or monovalent $Cu_2O$. While the tested Cu coupons have a native oxide layer, this is expected to be a mix of the two states. Oxidized Cu can be prepared in either state by annealing Cu coupons in air as described previously.[55] Coupons were prepared with CuO and $Cu_2O$ oxidized surfaces. Contact assays performed on oxidized coupons of CuO, Figure S3, showed comparable bioactivity to pristine copper, with log reductions of 6.5 and 4.8 for Phi6 and MS2, and no *E. coli* or *B. subtilis* were recovered. The $Cu_2O$ films were still bioactive, but showed more microbe-specific results, with log reductions of 2.8, 4.9, 5.8, and 4.5 for Phi6, MS2, *E. coli* and *B. subtilis*, respectively. These results suggest that the divalent ($Cu^{+2}$) oxide possesses stronger anti-microbial properties than the monovalent ($Cu^{+1}$).

*Candidate Prototype Currency*

In the final preliminary investigation, the bioactivity of Cu-containing alloy ($Cu_{55}Zn_{28}Ni_{14}Mn_2$) was evaluated. This alloy is a candidate material for next-generation minted currency,[56] making it highly relevant as a high-traffic touch surface. Current coins in circulation in the United States are alloys of Ni (8.3 % or 25%) and Cu, or are Cu-coated in the case of the penny. Contact assays showed significant ($p<0.05$) activity against Phi6 (log reduction of 3.2) and *E. coli* (4.5) (Supplemental Figure 4). However, the bioactivity against MS2 (2.0) and *B. subtilis* (0.5) was statistically identical to the 304 stainless steel control. This alloy is similar in composition to brass alloy 360, however, its bioactivity is significantly less, suggesting the Cu-content is not the only motivator of activity.

*CuNiCoAg Alloy Films*

A thin film comprised of the bioactive metals Cu, Ag, Co and Ni was prepared by sputtering and segmented into eight coupons, identified in Figure 5a, for testing in contact assays. As noted above, the rapid transformation from the gas phase to a solid film which is achieved with the sputtering technique prepares the as-grown film as a homogenized alloy.[37, 38] The bioactivity of each coupon was used in the preparation of a heatmap, shown for each organism in Figure 5, with the quantitative bar-graph representation shown in the Supplemental Figure 5. For Phi6, treatments with chips 2 and 5 resulted in the largest reductions of activity ($p < 0.05$), nearly at the limit of detection and comparable to pure Cu. Bioactivity sequentially decreased in coupons 6, 3, 8, 4, 7, and 1. Comparing Figure 5a with Figure 1a, the bioactivity against Phi6 closely follows the Cu content. Testing against the phage MS2 (Fig. 5b) had a similar trend, with coupons 5, 2, and 3 showing the strongest activity (defined as comparable to pure Cu) and decreasing with the Cu content. The coupons showed broader activity against *E. coli* and *B. subtilis*, Figures 5c and d, respectively, with coupons 2, 5, and 6 showing strong bioactivity. For *E. coli* the bioactivity continued to show a direct dependence on the Cu content, similar to Phi6 and MS2, however, the heat map suggests that the most-active composition against *B. subtilis* is near the equacompositional point.

Across all of the testing, coupon #1 ($Cu_{23}Co_{40}Ni_{25}Ag_{12}$) showed the least antimicrobial activity. This is notable because the heat plots tend to indicate a strong dependence on Cu content, however coupon #1 was not the lowest Cu composition (23 %); coupon #1 has the largest Co content (40 %). This result again suggests that Cu levels are not completely driving the alloy's activity. Finally, none of the plots show any sensitivity of the structural distribution shown in Figure 1c.

*CuNiCoAgSi Annealed Films*

The thin films were annealed, resulting in the phase separation discussed above and in Figure 3, then tested for anti-microbial activity. Testing results are shown in heat maps, Figures 5e-h, and in bar graphs in the (Supplemental Figure 6). Testing on Phi6, Figure 5e, again showed strong activity from coupon 2 which was statistically identical to the as-grown coupon 2. However, coupon 5 had a statistically significant reduction in bioactivity against Phi6 ($p<0.05$) compared to its as-grown alloyed counterpart. All of the other chips showed increased activity after annealing, however only chip 7 showed a statistically significant increase ($p<0.05$).

The heatmaps for MS2 and *E. coli* show activity trends which are similar between the as-grown and annealed samples. Against MS2 the annealed wafer shows slightly increased activity, while *E. coli* shows a slight decrease. However, neither trend showed a statistically significant difference between the as-grown and annealed test surfaces.

Finally, *B. subtilis* was tested and the coupons 2 and 7 again showed significant activity (Fig. 4h and S6). However, in contrast to the other pathogens, the other coupons showed a large reduction in activity relative to the alloyed sample. Coupons 1 and 3 still showed stronger activity compared to the SS control, while the remaining coupons (8, 6, 7, and 4) were statistically identical to the SS control. The annealing process led to a significant loss of antimicrobial activity on the chips 3, 6, 4, and 7 ($p<0.05$), which were from the Ni and Co heavy regions, when compared to the as-grown chips.

*Collective Bioactivity*

The above data shows the bioactivity of contact surfaces for a variety of common engineering materials and an alloyed and phase-separated thin-film. Four metals were chosen based off their previously reported bioactivity (copper, cobalt, nickel, and silver) and prepared as an alloyed thin-film which were then annealed, resulting in Si uptake and phase separation. While the materials design paradigm proposed to realize an alloy, stabilized by its chemical entropy, the phase separation shows that this was unsuccessful, the annealing resulted in phase separation. However, this allowed testing of the role of microstructure and chemical distribution. Contact viability assays were performed on two surrogate bacteriophages and two bacteria, representative of common human pathogens.

The prevailing theme throughout the measurement was the overwhelming bioactivity of copper against every surrogate organism. On pure Cu, Phi6, the SARS-CoV-2 surrogate was inactivated by nearly 7 orders of magnitude to after only 30 seconds. Each of the other three microorganisms were also significantly reduced to the limit of detection after < 30 minutes. The strong bioactivity of Cu is consistent with previous reports.[26, 28] Several mechanism of action have been proposed to support the bioactivity of Cu, including membrane disruption, enzyme inactivation, the generation of reactive oxygen species, and the denaturing genetic material. In comparison, the non-copper containing metals and oxides individually reduced all of the microorganisms' activity by approximately a single order of magnitude. These results were unexpected due to the well-known bioactivity of Zn and Ag in-particular. This may be the result of the microstructure which appears to be related to the bioactivity, especially for Ag. These metals were not treated before testing, and so likely have a native oxide layer, which may reduce their activity. However, much of the reported bioactivity emphasizes the role of the cation, which typically has an oxidation state of +2 and would be found in many of these native oxide surfaces.

The bioactivity of Cu and its two common oxides, CuO and $Cu_2O$, representing divalent and monovalent Cu, respectively, were tested. Both oxides showed strong activity, with the monovalent $Cu_2O$ being slightly less effective against Phi6 ($p<0.05$). The distinction between the two copper oxides is the valence state of the copper, $Cu^{2+}$ for CuO and $Cu^{1+}$ for $Cu_2O$. Previous works have emphasized the critical role of Cu ions in achieving bioactivity against microorganisms. For example, in viral assays, $Cu^{1+}$ ions were responsible for generating hydroxyl radicals that led to the deactivation of the influenza virus[26] and $Cu^{2+}$ showed activity against enveloped and non-enveloped viruses[57]. While in bacterial systems, $Cu^{1+}$ is reported to be more toxic to bacteria[26] and can initiate redox cycling which can damage key cellular processes. Due to the short testing time frame it is unlikely that the microbes caused surface corrosion, particularly since there was no observed drop in activity of the metal controls.[58,59] Next, two brasses were tested: alloy 360 ($Cu_{0.6}Zn_{0.4}$) and alloy 260 ($Cu_{0.7}Zn_{0.3}$). Alloy 260, which has higher Cu content, had stronger bioactivity and statistically identical within our assay to pure Cu. Finally, a currency prototype which is similar to brass, $Cu_{55}Zn_{28}Ni_{14}Mn_2$, showed moderate anti-microbial activity against Phi6 and *E. coli*, but weak activity against MS2 and *B. subtilis*. This was surprising because the currency has a similar composition as brass alloy 360, but replaces 14% of the Zn with Ni. The Ni is included for its anticorrosion qualities, supporting the idea that the oxidized surface is important to achieve bioactivity.

Considering the bioactivity of the as-prepared (alloyed) thin-film, the data suggests that, for all the organisms, the Cu content is a primary factor in the bioactivity. The bioactivity of the alloyed wafer against Phi6 (Figure 5a) is plotted going away from the Cu target across the equatorial ray of the wafer in Figure 6a. This data is expected to be sigmoidal, with both high- and low-concentrations of Cu having little dependence on the at.% Cu. Indeed, the data was well fitted with a variety of sigmoidal functions, with a 5-parameter Logistic function shown in the figure. Notably, the region 10 at.% - 65 at.%, which represent the composition explored on the wafer, is highly linear; a fit to this data returns a line: $Log(A) = 0.094\ x + 1.04$, where $A$ is the total bioactivity and $x$ is the at.% Cu. This equation can be arranged to solve for the total activity: $A = A_0 e^{cx}$, where the intercept ($x=0$) corresponds to the inherent activity at 0% Cu ($A_0=11$), and $c$ is a copper-derived bioactivity correlation constant (0.094). The value for $A_0$ agrees well with the activity of the stainless steel, which showed a reduction of 9.6 over the same time frame. Agreement in these values supports the relative inactivity of stainless steel.

Another important feature of the alloyed data is that, within the investigated compositional range (10 at.% - 65 at.%), the data does not show any flattening at high or low concentrations. This is notable because it implies that, even in relatively dilute concentrations the copper still contributes to the bioactivity. In other words, the critical density of copper to achieve some bioactivity is less-than 10 at.% – the lowest concentration tested here.

While the alloyed wafer shows a clear dependence on the Cu content, the annealed sample, also plotted in Figure 6a, is much less clear. The ray for the annealed sample shows an appreciable drop in activity for the high-concentrations and increased activity for the low-concentrations. This difference can be understood by considering the phase separation reported above. Specifically, the EXD images (Figure 3) showed that the Cu is concentrated into specific chemical domains of the film, resulting in Cu-rich and Cu-poor domains. As a result, the effective areal coverage of the copper on the film goes down. However, the Cu-rich domains will show exponentially increased activity, as shown from the alloyed samples. Since the bioactivity depends exponentially on the Cu content, while the areal coverage changes linearly, the resulting bioactivity in the annealed samples is increased. However, at higher concentrations of Cu, the bioactivity is already exceedingly strong, such-that additional Cu concentration does not increase activity. That is, the sigmoidal fit shows that increasing the Cu concentration beyond 65 at.% does not significantly increase the bioactivity on this timeframe – the phages are already inactivated. Therefore, the phase separation contributes only by decreasing the areal coverage and hence reduces the apparent bioactivity.

The other organisms are shown in Figure 5b-d. Of these, MS2 shows a similar, albeit weaker, trend to Phi6, with decreased (increased) activity after annealing at higher (lower) Cu concentrations. *E. coli* showed a slight (not statistically significant) increase in activity in the annealed sample, but in general the trend changed very little. Finally, *B. subtilis* showed a significant decrease in activity with annealing. One possible reason for this decrease in activity is the large size of *B. subtilis*, which is 4-10 μm in length and 0.5 μm in diameter, compared to *E. coli*, which is 1-2 μm in length and 0.5 μm in diameter. Both of these length scales are larger than the chemical domains (≈700 nm) after phase separation, however the *B. subtilis* may be large enough that it is able to isolate the copper damaged regions. Important to note, in this plot, at high concentrations of Cu, the wafer is able to fully kill the bacteria.

Figure 6 also shows that the testing time for each organism (30 s for Phi6, 30 min for MS2, 20 min for *E. coli*, and 10 min for *B. subtilis*) captures a range of efficacies, including fully sanitizing. The significant differences between these testing times may highlight critical physiological differences between the microorganisms which determine their susceptibility to the metallic surface. Specifically, Phi6, the enveloped phage, had the shortest testing time and is thus most susceptible to attack by the Cu in the surface. This may indicate that Cu, or the native oxides, rapidly degrades the envelope membrane or easily passes through the membrane, resulting in inactivation. The much longer (60× longer) testing times for the MS2 suggests that the protein capsid which surrounds MS2 provides significant protection against the Cu bioactivity. Another key difference between Phi6 and MS2 is that Phi6 has spike proteins, which are responsible for binding to the bacterial host, while MS2 has no such structure. The Cu bioactivity may be achieved by degrading, or otherwise denaturing, the structure of the spike proteins, achieving rapid inactivation.

Comparing the bacteria, the test times are also much longer than Phi6, but this may simply be a consequence of their much larger size. Between *E. coli* and *B. subtilis*, the shorter testing time of *B. subtilis*, which is also a larger bacteria, suggests it is more susceptible to the surface bioactivity. This is somewhat surprising because *B. subtilis* is the gram-positive surrogate, meaning it has an exterior cell wall, comprised of peptidoglycan enmeshed with wall teichoic and lipoteichoic acids, which collectively from sheets of anionic charge. These layers act as a semi-permeable barrier that regulates cation movement, through binding, and is expected to protect the more-fragile membrane from the reactive metallic surface.[60] Further studies beyond the scope of this work are necessary to elucidate the various modes of action in these systems.

**Discussion**

In summary, this work investigated a range of materials as candidate anti-microbial, high-traffic touch surfaces.[61] Four specific metals (Co, Ni, Cu and Ag) were chosen based off previous literature, and prepared as a thin-film medium-entropy alloy. Upon annealing, the alloy took up Si from the substrate and phase separated into Ag, Cu-rich and Cu-poor regions. The candidate materials were tested against four microorganisms which are surrogates for common viral and bacterial human pathogens. The collective results emphasize the overwhelming bioactivity from Cu. Key results were that (1) Cu was an effective bioactive agent even at low concentrations, (2) the inclusion of anti-corrosion elements tended to suppress bioactivity, and (3) the bioactivity was exponentially dependent on the Cu concentration. This last result was balanced against the areal density of Cu in the phase separated (annealed) films, increasing the bioactivity without adding further Cu. Together, these results demonstrate copper containing alloys as effective antimicrobial touch surfaces and offers a variety of insights supporting the development of effective sanitizing surfaces.

**Materials and Methods**
*Purchased Metal Materials*
Some single-element and binary alloys were purchased from McMaster Carr, including Sn, Cu, SS304, Brass 360 (*$Cu_{60}Zn_{40}$*) and Brass 260 ($Cu_{70}Zn_{30}$); these were industry grade materials. Silver plate (99.9%) was purchased from Sigma Aldrich. The Si sample was a semiconductor-grade (100) single crystal wafer. $SiO_2$ was a fused silica microscope slide. All materials were used as received without any surface treatments. Copper oxides (I) and (II) were prepared by annealing plate copper in air at 300 °C and 350 °C, respectively, for 3 hours, then passively cooling to room temperature, also in air.

*Quaternary Alloys*

Combinatorial wafers were prepared by magnetron sputtering in an Ar atmosphere from single-element sources, confocally oriented around an (001) Si wafer with native oxide coating. Deposition was performed at room temperature. The wafer was not rotated during deposition, resulting in compositional gradients across the sample with a target film thickness of ≈200 nm. Compositions were controlled by adjusting the power to the sputtering sources, resulting in an approximately equacompositional point located at the center of the wafer. Precise measurement of the compositions was determined using electron dispersive X-ray spectroscopy (EDX) at 17 points along eight directions oriented radially from the wafer center. These measurements also captured element-specific plane-view maps of the local compositional distribution. Four as-grown wafer alloys were prepared, one for each surrogate microorganisms. Annealed samples were then heated in vacuum (<$10^{-6}$ Torr) or forming gas (4% $H_2$ in $N_2$) to 600 °C and held for 1 hour, then allowed to passively cool overnight.

*X-ray Diffraction*

X-ray diffraction (XRD) measurements were performed using a Cu K-α ($\lambda$=1.5406 Å) source. The sample rotation axis ($\theta$) was fixed and the sample illuminated with a spot-like beam, as the $2\theta$ axis was swept from 20°-80°. The sample was then translated, and the measurement repeated, capturing the XRD pattern from 17 points, approximately coinciding with the EDX measurements.

*Scanning Electron Microscopy*

Scanning electron microscopy (SEM) was performed on the as-grown and annealed samples using an acceleration voltage of 15 keV and a backscattered electron detector. Energy dispersive X-ray spectroscopy (EDX), including the spectrum and spatial mapping, was performed using a cryogen-free silicon drift detector.

*Currency Prototype*

The prototype currency composition was 14.4 wt. % Ni, 28.1 wt. % Zn, 2.3 wt. % Mn, balance Cu. Only a single coin was available for contact testing. The initial purities of the pure element starting materials were 0.9995+ Cu, Ni, and Zn, and 0.99+ Mn. Sheets of the commercial alloys C71300 (Cu-25Ni) and C77000 (Cu-18Ni27Ni)3 were provided by the US Mint for characterization and comparison with the designed prototype alloys. Coins were prepared as described by Lass et. al., 2018[56].

*Bacterial Strains and Bacteriophages*

The bacteria P*seudomonas Syringae var. phaseolicola* (Félix d'Hérelle Reference Center for Bacterial Viruses, University of Laval, QC Canada) was used for phage titering and phage propagation of *Psuedomonas* phage Phi6. *Escherichia Coli* C3000 (ATCC 15597, American Type Culture Collection, VA USA) was used for phage titering and phage propagation of the *E.coli* phage MS2 (ATCC 15597-B1). *Escherichia coli* FDA strain Seattle 1946 (ATCC 25922) was used as the model gram negative strain for bactericidal testing. *Bacillus subtilis subsp. spizizenii* NRS 231 (ATCC 6633) was used as the model gram positive strain for bactericidal testing. All strains were stored at -80°C in Brain Heart Infusion (BHI) supplemented with 15 % (wt/vol) glycerol. Cultures of *P. syringae* were grown on Lysogeny Broth (LB) agar plates at 25 °C. Overnight cultures were inoculated with a single colony from a streak plate into LB broth and grown at 25 °C in a shaking water bath at 160 RPM. Both *E. coli* strains were grown on LB agar plates at 37 °C, and overnight cultures were prepared as previously described and incubated at 37 °C. *B. subtilis* cultures were grown on BHI agar plates at 37 °C, and overnight cultures were prepared in BHI media and grown at 37 °C.

Phage stocks for this project were amplified by the plate lysate method. An aliquot of overnight culture (30 µL for MS2 and 300 µL for Phi6) and 100 µL of the phage dilution in pH 7.4 phosphate buffered saline (PBS) was aliquoted into 3–3.5 mL of LB overlay agar after equilibrating to 45 °C. The mixture was vortexed briefly, poured onto the agar underlay and allowed to solidify for 20–30 min. Plates were then incubated for 18–24 hr at 25 °C for Phi6 or 37 °C for MS2. Five mL of sterile PBS was aliquoted onto each plate with confluent lysis and allowed to sit for 1–2 h. The buffer was then removed from the plate with a serological pipette, centrifuged at 5000 g for 10 min at 4 °C to remove debris, and then filter sterilized using a 0.20 µm-pore size surfactant-free cellulose acetate (SFCA) syringe filter (Corning, Incorporated, Corning, NY). This filter sterilized stock was then concentrated by centrifugation at 12,000 g for 2 hr, the supernatant was removed by serological pipet and the phage pellet was resuspended in sterile PBS by static incubation overnight at 4 °C. Master stocks of Phi6 were stored at -80 °C in PBS supplemented with 15 % (v/v) glycerol, working stocks were prepared by plate lysate and centrifuge concentration from freshly thawed master stock prior to each experiment and stored short-term at 4 °C. All stocks of MS2 were stored at 4 °C.

Bacterial stocks for bactericidal testing were prepared by inoculating a 5 mL tube of the appropriate growth media 1:100 with an overnight culture of the test bacteria and incubated in a shaking water bath at the 37 °C at 160 RPM until the $OD_{600}$ reached 0.2. The culture was then centrifuged at 5000 g for 10 minutes, the supernatant was discarded and the bacterial pellet was resuspended in 500 µL of PBS and immediately used for testing.

*Bactericidal and Virucidal testing.*

Bioactivity was determined through contact assay testing. All preparation and testing took place in a class II biosafety cabinet. Prior to exposure all test materials were briefly washed in acetone, and once dried each material was placed into a 60 x 15 mm sterile petri dish. Each microorganism was exposed to fresh (untested) coupons of Cu, Ag, Ni, Co, oxides and brass ($2.5 \times 2.5$ cm$^2$). The alloy wafers were segmented into 8 chips that were $2.5 \times 2.5$ cm$^2$ in size and each microorganism was tested on a single set of chips. All contact assays were conducted in triplicate and on separate occasions. To conduct the assay 10 μL of a working stock of the test microbe was spotted on to center of the test material, covered by a $25 \times 25$ mm$^2$ square glass slide and incubated at room temperature for the designated exposure time: Phi6 30 sec, MS2 for 30 min, *E. coli* for 20 min, and *B. subtilis* for 10 min. After the exposure time had elapsed, the slide was separated from the test material by 70% ethanol-decontaminated forceps, and both the test material and slide were washed three times with a single 990 μL aliquot of PBS into the petri dish. The wash was immediately collected into a sterile 1.5 mL microcentrifuge tube, and then serially diluted in PBS and enumerated by double-layer agar plating for phages and by spread plating for bacteria. After testing, all materials were immediately decontaminated in 70% ethanol for 30 min, then rinsed with deionized water, dried, and rinsed with acetone before storage under vacuum.

*Statistical Analysis of Microbial Contact Assays.*

To evaluate the efficacy of each metal treatment the log reduction was calculated from the raw phage and microbial counts. A Shapiro-Wilks test was conducted and if the data did not meet the assumptions of normality, a rank transformation was applied. A one-way analysis of variance (ANOVA) ($p < 0.05$) was conducted on the data (normal and transformed) to determine any significant differences with the statistical software JMP Pro 14.2 (SAS Institute, Cary, NC).

## ACKNOWLEDGEMENTS

This research was supported by the Science Alliance at the University of Tennessee, Knoxville, thorough the JDRD Collaborative Cohort Program Fellowship. Bioactivity testing was partly supported by NSF award 2028542.

## AUTHOR CONTRIBUTIONS

Du.A.G., T.D., A.F.M. Y.Y. and P.D.R. conceived and designed the experiments. A.F.M., D.B., and T.D. conducted the microbial experiments. Da.A.G. and P.D.R. fabricated the wafers and E.L. fabricated the currency prototype. Da.A.G., C.S.J, W.L. and N.T performed the materials characterization. A.F.M., Du.A.G., and T.D wrote the manuscript. All authors contributed to the discussion and manuscript revision.

**Tables**

| # | Cu | Co | Ni | Ag |
|---|----|----|----|-----|
| 1 | 23 | 40 | 25 | 12 |
| 2 | 44 | 22 | 15 | 19 |
| 3 | 25 | 24.5 | 28.5 | 22 |
| 4 | 15 | 20 | 46 | 19 |
| 5 | 42 | 12 | 15 | 31 |
| 6 | 25 | 11 | 28 | 35.6 |
| 7 | 14 | 11.5 | 43 | 31.5 |
| 8 | 19 | 6 | 21 | 54 |

**Table 1.** Percentage (%) of metal composition at the center of each coupon used for bioactivity testing.

**Figures**

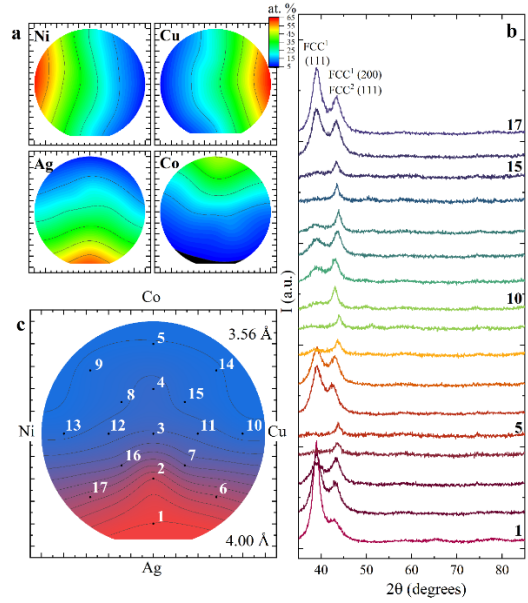

**Figure 1** (a) Compositional map of Ni, Cu, Ag and Co. (b) X-ray diffraction patterns, taken at positions 1-17, identified in panel c. (c) Contour plot of the two FCC phases in the as-grown system.

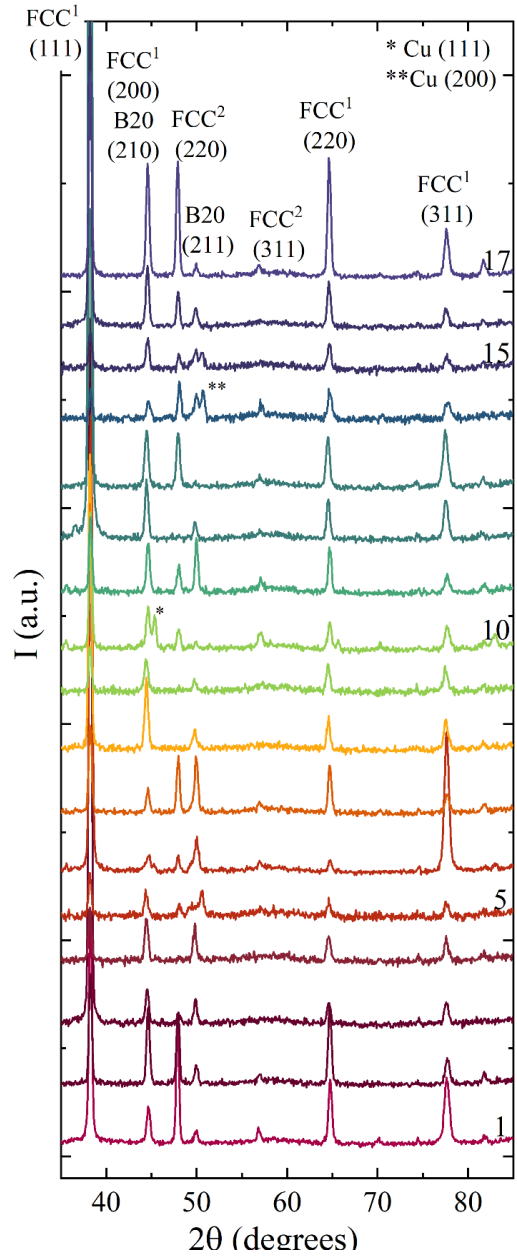

**Figure 2** X-ray diffraction patterns captured at positions 1-17 after the anneal.

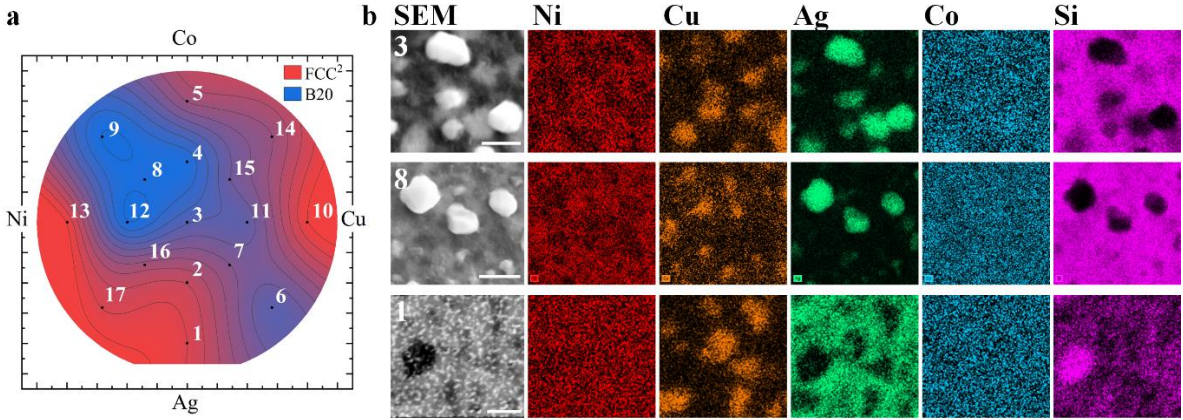

**Figure 3** (a) Contour plot showing the relative XRD peak intensity for the FCC phases after the anneal, $A(50°) / (A(47°) + A(50°))$, where A is the peak area. (b) SEM and EDX images of positions 3, 8, and 1, identified in the contour plot. The scale bar in the SEM image corresponds to 1000 nm.

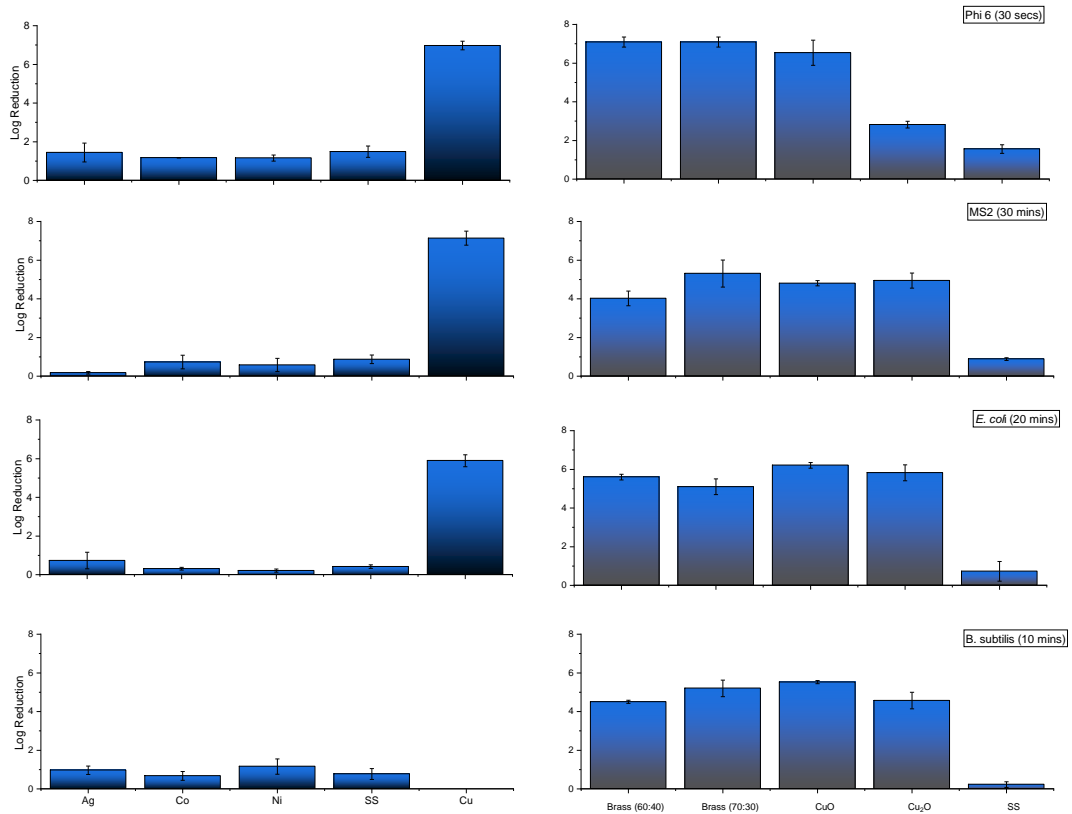

**Figure 4.** Bioactivity of single phase and binary alloys of (a) Phi6, (b) MS2, (c) *E. coli*, and (d) *B. subtilis*. No *B. subtilis* was recovered on the metal Cu.

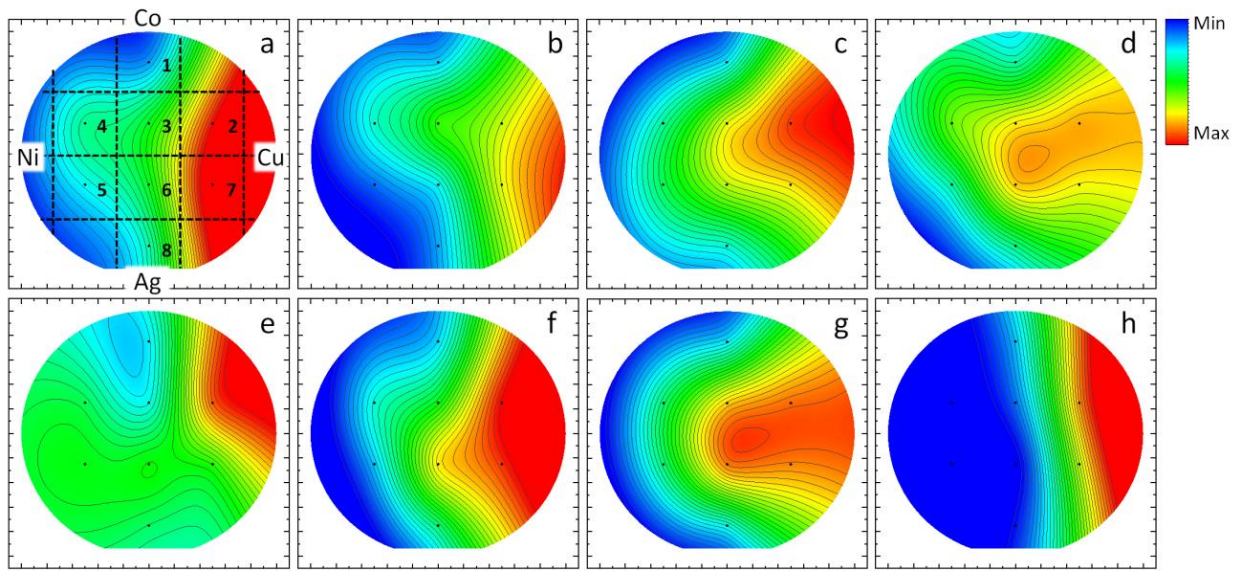

**Figure 5.** Heat map of log-reduction on the thin film chips of a) Phi6, b) MS2, c) *E. coli*, and d) *B. subtilis*; log-reduction on the annealed chips of e) Phi6, f) MS2, g) *E. coli*, and h) *B. subtilis*.

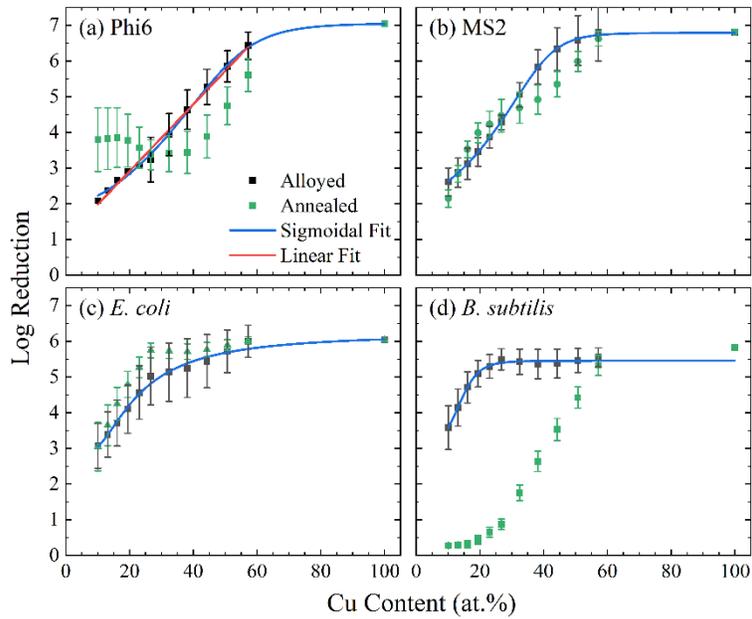

**Figure 6.** Line-cuts of the bioactivity for (a) Phi6, (b) MS2, (c) *E. coli*, and (d) *B. subtilis*, taken across the equatorial ray of the wafers between Cu and Ni.